\def\ArxivDraft{1}
\newif\ifshowauthors
\ifdefined\AuthorDraft\showauthorstrue\else\showauthorsfalse\fi
\ifdefined\ArxivDraft
  \showauthorstrue
  \documentclass[sigconf,nonacm]{acmart}
\else
  \ifshowauthors
    \documentclass[sigconf,review]{acmart}
  \else
    \documentclass[sigconf,review,anonymous]{acmart}
  \fi
\fi
\setkeys{acmart.cls}{balance=false}
\setcopyright{none}
\ifdefined\ArxivDraft
  \acmConference[Research preprint]{Research preprint}{November 2026}{}
\else
  \acmConference[Draft]{Unsubmitted research draft}{November 2026}{No venue assigned}
\fi
\acmDOI{}
\acmISBN{}
\usepackage{amsmath}
\usepackage{graphicx}
\usepackage{booktabs}
\newcommand{\sys}{\textsc{ckks-train}}
\newcommand{\rot}{\operatorname{rot}}
\newcommand{\KS}{\operatorname{KS}}
\newcommand{\ceil}[1]{\left\lceil #1\right\rceil}
\newcommand{\floor}[1]{\left\lfloor #1\right\rfloor}
\title[GPU-Accelerated Encrypted Training]{GPU-Accelerated Non-Interactive Encrypted Neural Network Training with Batched Diagonal Packing}
\ifshowauthors
  \author{Kentaro Mihara}
  \authornote{Independent Researcher.}
  \email{miha.ken.19@gmail.com}
\fi
\begin{document}
\begin{abstract}
Training a neural network under homomorphic encryption requires transposes, batch reductions, optimizer updates, and repeated refreshes in addition to forward matrix products. We combine rectangular diagonal packing, feature-major SIMD mini-batch lanes, baby-step giant-step evaluation, and compact optimizer state in an MIT-licensed Rust CKKS implementation. Its CUDA path keeps training and bootstrap ciphertext arithmetic on the GPU, while retaining host key generation and client operations. In a matched 100-update experiment on the main non-interactive MNIST model, device-resident bootstrapping reduces training time from 14,268.295 to 293.075 seconds, a $48.68\times$ ratio against our CPU implementation, with exactly equal decoded parameters and predictions. Including setup and final evaluation gives a $40.67\times$ ratio. A separate GPU run completes ten passes over 4,800 images in 2,967.095 seconds and achieves 90.35\% test accuracy versus 90.41\% for its plaintext twin. These are single-initialization results; exact CPU/GPU parameter agreement is verified at 100 updates, not for the full 1,000-update run. Earlier completed CPU runs support a larger-batch key-switch advantage and baseline-batch accuracy within the registered tolerance, but the baseline-batch normalized key-switch target is missed. The released ReBoot baseline does not meet the common security-estimator threshold, and its full-setting CPU time is unmeasured. We therefore distinguish measured acceleration of our implementation from conditional layout comparisons and do not claim unconditional superiority over ReBoot.
\end{abstract}
\maketitle
\section{Introduction}
\label{sec:intro}
Homomorphic encryption allows a service to evaluate a public computation without learning the encrypted inputs. Training is particularly demanding: the model changes at each step, gradients require transposed products, and a long optimization trajectory repeatedly exhausts the available multiplicative depth. A useful packing scheme must therefore account for the entire update, not just inference.

Mihara et al.~\cite{mihara2020} showed that diagonal-packed weights support transposition through rotations rather than ciphertext multiplications. Their interactive training design used a client to restore ciphertexts between updates. ReBoot~\cite{reboot2026} subsequently demonstrated non-interactive CKKS neural-network training, combining packed arithmetic, local learning signals, and bootstrapping. Non-interactivity itself is therefore not a novelty claim of this work. Our question is whether diagonal packing remains competitive when batch reductions, server-side layout maintenance, optimizer state, and bootstrapping are all charged.

We develop \sys{}, an MIT-licensed Rust implementation, and study three questions. \textbf{RQ1} asks how many key switches a complete training update requires compared with a ReBoot-style layout and the released ReBoot baseline. \textbf{RQ2} asks how bootstrap placement interacts with depth and compact state. \textbf{RQ3} asks what CUDA execution and its constituent optimizations contribute to measured training cost. A secondary continuity experiment revisits the original Iris setting; it is an interactive demonstration, not evidence of non-interactive training performance.

Our contributions are: (i) a rectangular, batched diagonal layout with a rotation-only transpose and explicit reduction costs; (ii) a compact-state non-interactive update whose operation counts can be checked against structural predictions; and (iii) a CUDA training path that keeps ciphertext arithmetic, including bootstrapping, on the device while preserving the CPU computation's operation order and numerical results in the measured comparisons. The third contribution is evaluated at the training level, charging repeated refreshes, input preparation and setup through explicitly separated timing boundaries. The MIT implementation and retained source, input, executable and result identities make those measurements inspectable. Hoisting, baby-step giant-step (BSGS), sparse-secret encapsulation (SSE), and bootstrapping are established techniques; the contribution lies in their training-specific composition and measured execution, not in inventing these primitives.

\paragraph{Status of the evidence.}
The generated tables include two completed full-setting encrypted runs: FULL-1 at the larger batch and FULL-2 at the released baseline batch, each with initial-weight seed zero. FULL-2 meets the registered accuracy tolerance, but misses the key-switch target under the conservative common-baseline-bootstrap convention. FULL-1 meets the count target under all four conventions; its batch differs from the accuracy-gate configuration. The released baseline and our preset also have different secret distributions and security estimates, and full-setting baseline CPU timing is unmeasured. The registered overall RQ1 goal is therefore unmet. We report a conditional comparison and re-evaluation rather than an equal-security superiority claim. The earlier encrypted subsets have a different input dimension. The matched main-model device-bootstrap experiment yields a $48.68\times$ CPU-over-CUDA training-time ratio over 100 updates, with exactly equal decoded parameters and predictions. A separate GPU run completes ten passes over 4,800 examples in 2,967.095 seconds with 90.35\% test accuracy. These single-initialization results do not establish superiority over an unmeasured ReBoot GPU baseline. Section~\ref{sec:limits} states these limits.

\section{Preliminaries and Threat Model}
\label{sec:prelim}
\paragraph{Arithmetic.}
CKKS~\cite{ckks2017} encodes approximate vectors in
$R_Q=\mathbb{Z}_Q[X]/(X^N+1)$, with up to $N/2$ complex slots. We use real slot values. A ciphertext at level $\ell$ uses $q_0,\ldots,q_\ell$ and a scale $\Delta$. Multiplication followed by rescaling consumes a modulus prime. Rotations permute slots through a Galois automorphism and a key switch. Hybrid RNS key switching uses special primes $P$ and a digit decomposition; its cost depends on the active limbs, digit sizes, and memory traffic.

We count rotations (including conjugations), relinearizations, and explicit SSE switches:
\begin{equation}
  \KS = n_{\mathrm{rot}}+n_{\mathrm{relin}}+n_{\mathrm{SSE}}.
  \label{eq:ks}
\end{equation}
A ciphertext tensor product is not automatically a key switch: lazy relinearization may combine several products before one switch. Bootstrap events are recorded separately and their constituent operations are already in Eq.~\eqref{eq:ks}; adding a bootstrap-equivalent charge again would double count. Reported counts cover the whole network and are averaged over processed examples, not over layers. A count is a workload metric, not a machine-independent latency estimate.

\paragraph{Refreshes.}
An interactive refresh decrypts and re-encrypts at the key owner. A bootstrap instead evaluates a refresh circuit using public evaluation keys. Our non-interactive path uses SSE~\cite{sse2022}, centered modulus raising, coefficient-to-slot and slot-to-coefficient transforms, and an approximate modular-reduction polynomial. Public transforms use collapsed stages and BSGS; exact scale tracking makes them composable with training.

\paragraph{Protected objects and public leakage.}
We consider a semi-honest evaluation server and a single trusted key owner, under the usual RLWE and evaluation-key assumptions of the chosen CKKS construction. No new security reduction is claimed. Training examples, labels, model weights, intermediate activations, gradients, and optimizer state are encrypted. The architecture, dimensions, batch schedule, hyperparameters, parameters, evaluation keys, and algorithmic control flow are public. The key owner initializes and encrypts the data and initial model, and decrypts the resulting model. Known initialization distributions are not themselves secret. This model does not hide dimensions, resource use, access schedules, or timing, and it does not provide malicious-server verification, differential privacy, multi-party key custody, or protection against inference from a legitimately decrypted model.

\paragraph{Diagnostics are not a server capability.}
The benchmark process can also instantiate a secret-key holder, decrypt checkpoints and predictions, compare against a plaintext twin, and report accuracy or activation ranges. Those are client-side diagnostics for a trusted experiment harness; they must not be exposed as server logs in a deployment. The generic non-interactive trainer has no secret key and makes public depth-based refresh decisions. Disabling diagnostic decryption changes neither the encrypted arithmetic nor the public schedule. The interactive path explicitly involves the client during training and is labeled separately. The in-library ReBoot-layout microbenchmark uses plaintext labels and client-side optimizer updates: it supports arithmetic/count comparisons, not the full confidentiality claim or a faithful non-interactive baseline.

\section{Rotation-Only Transposition for Mini-Batch Training}
\label{sec:packing}
\subsection{A common layout for activations and weights}
Let $B$ be a power-of-two lane count and $d$ a power-of-two feature capacity such that $dB$ divides the slot count. Slot $jB+b$ stores feature $j$ of sample $b$. Blocks repeat when $dB$ is smaller than the slot count. A layer has positive integer input and output dimensions $d_i,d_o$. Define the padded dimensions $n_i=2^{\ceil{\log_2 d_i}}$ and $n_o=2^{\ceil{\log_2 d_o}}$, and require $n_i,n_o\le d$. A vector of padded dimension $n\in\{n_i,n_o\}$ is canonical when its feature sequence is $n$-periodic in every lane; unused logical features are zero. This periodicity eliminates a separate replication step at tall layers.

Zero-pad the original $d_o\times d_i$ weight matrix to $n_o\times n_i$. Let $m=\min(n_o,n_i)$ and $L=\max(n_o,n_i)$. Both divide $d$, so the $L$-periodic diagonals, rotations modulo $LB$, and integer logarithmic folds below are well-defined. Store the generalized diagonals
\begin{equation}
 D_c[j]=W[j\bmod n_o,(j+c)\bmod n_i],\quad 0\le c<m.
 \label{eq:diagonal}
\end{equation}
Weights are shared across all lanes. With BSGS split $m=gu$ and $c=ga+b$, the stored diagonal is
$P_c=\rot(D_c,-gaB)$. The encrypted product is
\begin{equation}
 y=\sum_{a<u}\rot\!\left(\sum_{b<g}P_{ga+b}\odot\rot(x,bB),gaB\right).
 \label{eq:bsgs}
\end{equation}
Here a positive rotation reads slot $s+k$ into output slot $s$; $\odot$ denotes a slot-wise product. For a wide layer, $\log_2(n_i/n_o)$ rotations and additions fold partial outputs together. For a tall layer, the canonical input already has the required replication. Neither case adds a mask to the fold.

\subsection{Transpose without products or masks}
Rectangular diagonal transposition without ciphertext multiplications is inherited from Mihara et al.~\cite{mihara2020}. Their general method uses rotations and additions; their implementation also reports OpenMP batch processing. Here we compose that idea with the periodic $m$-diagonal representation, feature-major SIMD lanes, and BSGS pre-rotations.

The transpose diagonal $E_i[j]=W[(j+i)\bmod n_o,j\bmod n_i]$ equals a rotation of one $D_c$. Choose $c=(-i)\bmod m$, and set $t=-c$ for a wide layer or $t=i$ for a tall layer. In stored BSGS form,
\begin{equation}
 Q_i=\rot(P_c,\,[t-g\floor{i/g}+g\floor{c/g}]B).
 \label{eq:transpose}
\end{equation}
The rotation amount is taken modulo $LB$. Thus transposition uses neither a ciphertext product nor a mask and consumes no multiplicative level. It is \emph{not} free: each nonzero rotation requires a key switch. Dedicated keys minimize rotation count; composing existing baby/giant keys saves key memory at the expense of extra switches. Our training runs use key reuse, and the extra work is included in the counters.

\subsection{Gradients, reductions, and a symbolic cost model}
For layer gradient $\delta$ and input $a$, a pre-rotated weight-gradient diagonal is
\begin{equation}
 \nabla P_{ga+b}
 =\operatorname{batchsum}\bigl(\rot(\delta,-gaB)\odot\rot(a,bB)\bigr).
 \label{eq:gradient}
\end{equation}
Batch summation first reduces lanes by powers-of-two rotations. A lane-zero mask isolates the reduced value; inverse-direction rotations replicate it when replicated gradients are required. The mask has nonintegral polynomial coefficients in the CKKS embedding, so encoding it at unit scale is not an exact, zero-depth shortcut. It consumes a level. Constants for the learning rate and batch normalization are folded into a necessary plaintext multiplication at its exact target scale.

\begin{table}[t]
\centering\small
\caption{Symbolic layer costs for the implemented layout, before any wide-layer fold. $g,u$ are the BSGS split, $m=gu$. Gradient costs show the replicated interactive form; compact non-interactive state changes the final replication.}
\label{tab:symbolic}
\begin{tabular}{lcc}
\toprule
Operation & Rotations & Relins \\
\midrule
BSGS encrypted matvec & $g+u-2$ & $u$ \\
Dedicated transpose & $m-1$ & $0$ \\
Replicated weight gradient & $g+u-2+2m\log_2 B$ & $m$ \\
\bottomrule
\end{tabular}
\end{table}
Hoisting reuses decomposition for rotations of the same input~\cite{halevi2018}. It changes wall-clock work but not Eq.~\eqref{eq:ks}. Lazy relinearization applies within a matvec giant-step sum. It cannot remove the separate relinearizations needed before gradient-diagonal rotations. These distinctions matter when interpreting ablation results.

\paragraph{Extensions relative to the interactive design.}
The first two extensions studied here are the periodic $\min(n_o,n_i)$-diagonal representation with feature-major SIMD mini-batch lanes, and its composition with hoisting and BSGS. General rectangular transposition and batch processing themselves are inherited capabilities, not novelty claims. The remaining four extensions---bootstrap placement, polynomial activations, momentum, and loss conventions---are specified in Section~\ref{sec:noninteractive}. None removes the cost of maintaining canonical server-side state; the initial interactive cost model omitted some replication work and was therefore too optimistic.

\section{Non-Interactive Training}
\label{sec:noninteractive}
\subsection{Compact state and the complete update}
Refreshing every replicated diagonal separately wastes bootstrap slots. We instead reduce each gradient to one lane per feature and combine masking, scaling, and lane shifting in a plaintext product. Selected lanes from several diagonals form compact weight and momentum containers. A public compact-layout plan fixes their offsets and the required rotations; unpacking restores the periodic diagonals for the next forward pass. Bias slots can share unused compact positions. Masked packing and unpacking both have a depth cost and are included in the closed-form predictor.

The server performs encrypted forward propagation, forms the loss gradient from encrypted labels, propagates through the transposed weights, and accumulates gradients over micro-batches before one optimizer update. Increasing the effective batch beyond the physical lane count therefore amortizes optimizer refreshes without pretending the examples fit in one ciphertext. It also changes optimization behavior and cannot be treated as a cost-only substitution.

The original dense-secret bootstrap constants did not satisfy the desired precision and depth budget. The implemented preset instead uses a dense ternary main key and a temporary sparse key during modulus raising. The SSE pair is returned to the dense main key before the remaining arithmetic. Collapsed transforms and a level-optimal Chebyshev evaluator fit the selected modulus bands. This is an engineering parameter choice, not a new bootstrap construction or an inherited failure-probability theorem. Appendix~\ref{app:params} specifies the chain and the exposed key instances.

\subsection{Three refresh schedules (extension 3)}
\texttt{AfterUpdate} (AU) bootstraps every compact weight and velocity container after each update. \texttt{WhenNeeded} (WN), an additional schedule beyond the original plan, does so only if the next update would exceed the remaining depth. The decision uses the next batch's fullness: padded lanes at a linear output need a validity mask, so a partial batch can consume an extra level. An unknown next shape is conservatively treated as partial.

\texttt{AfterActivation} (AA) refreshes each hidden activation and retains the WN rule for compact weights. For the shallow eMLP-1 workload, activation refresh does not reduce the frequency of weight refresh beyond WN; it adds a bootstrap per hidden activation and micro-batch. Its motivation is enabling deeper circuits that otherwise exceed the budget, not an automatic speedup. Activation magnitudes must fit the supported bootstrap range. The observed AA subset run exceeded the conservative range in some slots and had a larger activation error; it was not clipped, and does not certify arbitrary activation distributions.

For a sequence of identical updates consuming $D$ levels with $L_a$ usable levels after unpacking, $\floor{L_a/D}$ is only an initial scheduling bound. Unpack costs between updates, partial batches, and the actual state level must also enter the scheduler. We check the structural predictor against every measured update rather than replacing this stateful schedule with a fixed average.

\subsection{Activation, optimizer, and loss (extensions 4--6)}
\paragraph{Polynomial activation.}
We retain the earlier squared activation for continuity and implement $z^2+z$ for the ReBoot comparison. A polynomial quartic option is available for depth experiments. Its additional depth and available tests do not establish a measured accuracy improvement; that planned claim remains unsupported.

\paragraph{Momentum.}
Both SGD and Nesterov-style momentum are supported. Writing $h=\eta(1+\mu)G/n$ and scaled velocity $U=\eta(1+\mu)V$ gives
\begin{align}
 U'&=\mu U+h,\nonumber\\
 W'&=W-h-\frac{\mu^2}{1+\mu}U.
 \label{eq:nag}
\end{align}
Constants are folded into exact-scale products already needed by the gradient path. A changing learning rate requires rescaling the stored velocity representation, and its lane shifts are charged. Biases use SGD; the faithful ReBoot configuration freezes their encrypted zero values while retaining the implementation's bias arithmetic. The encrypted trainer evaluated here omits weight decay; the captured faithful setting uses zero decay. The separate plaintext paper-setting reconstruction includes weight decay and is labeled separately in the evaluation.

\paragraph{Loss.}
A squared-error loss yields an affine output gradient; MSE normalization and ReBoot's sum reduction are matched through the learning-rate convention. The faithful configuration uses the released initialization, sample order, output padding, and schedule, not just the paper's nominal hyperparameter names. Softmax/cross-entropy approximation is optional in the plan and is not implemented or evaluated here. Block MLPs follow local classifier losses and stop gradients across blocks; they must not be described as an ordinary deeper back-propagation network.

\section{GPU Implementation}
\label{sec:gpu}
\sys{} separates RNS arithmetic, CKKS evaluation, neural-network operations, CUDA execution, and benchmark orchestration. Generic \texttt{HeOps} and \texttt{BootstrapOps} interfaces let the same trainer use CPU or device ciphertexts. The evaluator checks parameter identity, levels, scales, key availability, and device ownership before work. Counters commit after success; the CUDA bootstrap also waits for final device synchronization before merging its internal counters.

The CUDA backend uses device NTTs, basis conversion, pointwise arithmetic, hybrid key switching, and BSGS matvecs. Small public launch data use pinned transfer buffers. Reusing a hoisted decomposition avoids repeating the expensive preparation for related rotations. Device/CPU bit parity is meaningful only for cases actually executed; memory-skipped test functions can still report process success.

\paragraph{Streams and memory.}
Independent products may be issued to several streams, but a training step's dependent matvecs are executed sequentially. The measured pre-fix multi-stream benchmark was slower than sequential execution (Table~\ref{tab:gpu_vs_cpu_ops}). Calling-thread submission and pool handling were subsequently revised. The remeasurement improves the latency of four independent products at a capped level under each tested pool threshold, with the preferred stream count depending on the threshold. This does not identify the contribution of each code change or establish a training-step speedup. An effective-bandwidth estimate for addition does not prove matvec occupancy or full-device saturation. A separate profiled run completed, but its trace has not been analyzed; that explanation remains a hypothesis.

Training uploads its Galois keys only through the highest level needed by the training circuit. For the main non-interactive workload this is level~16, while bootstrapping requires a level-31 relinearization key. Uploading that key separately avoids expanding all 35 training Galois keys to the bootstrap level. Bootstrap keys and transform diagonals are uploaded once. This preserves the original cryptographic parameters and operation order while fitting the evaluated workload on a 24~GiB RTX~4090. Allocation preflight combines the existing training estimate with a separate conservative 20~GiB free-memory floor for device bootstrap. Pool trims occur after large setup stages and at epoch boundaries. Estimates, admission floors and pool checkpoints are not measured memory peaks, and preflight does not guarantee later allocations under contention.

\paragraph{Bootstrap boundary.}
The CUDA bootstrap keeps ciphertext arithmetic on the device through SSE, modulus raising, transforms, and EvalMod. A narrow shared polynomial executor preserves the CPU operation order, integer rounding and scale labels. Public integer constants are still materialized on the host and uploaded; measured refresh time includes that overhead and final device synchronization. An executed full-slot $N=2^{16}$ gate on RTX~4090 matched every CPU ciphertext limb and operation counter, measured 17.4775 bits of precision over all 32,768 random real slots, and took 473.068~ms for one refresh. This misses the fixed 220~ms target. The observation is a single timed refresh, not a latency distribution or a peak-memory measurement.

The main benchmark selects this path with \texttt{--device-bootstrap}. The same generic trainer then executes its ciphertext operations, including non-interactive refresh, on CUDA without a secret-key decrypt/re-encrypt step. The historical CPU bootstrap adapter remains available for reproduction; interactive training is a separate client-refresh mode. Key generation, input encryption, public-constant preparation and final diagnostic decryption remain host work. We therefore report synchronized updates, training totals and wall time including setup separately, rather than describing the entire application as GPU-only.

\section{Evaluation}
\label{sec:evaluation}
\subsection{Protocol and evidence classes}
We distinguish a released baseline execution, an in-library layout microbenchmark, and a cost model. The first supplies observed wall-clock and bootstrap events; the second isolates arithmetic layout counts under our evaluator; the third supplies uninstrumented baseline quantities. A counter-derived normalization is not a separate execution. Historical generated tables retain the working name \emph{hekit} for \sys{} and its crate names.

The headline uses MNIST reduced to $196$ inputs and the complete released encrypted-comparison setting: $4800$ training examples over ten passes, a $196\!\to\!32\!\to\!16$ eMLP-1 with $z^2+z$ hidden activation, frozen zero biases, sum-loss NAG with momentum $0.9$, and the released sample order and learning-rate schedule. This is distinct from the $60000$-example paper-setting reconstruction. FULL-1 at effective batch $64$ and FULL-2 at the released batch $48$ in $64$ physical lanes have both completed with initial-weight seed zero. They process the same ten-pass sample count with different update counts. The command's single epoch traverses flattened ten-pass indices; its measured total is not one conventional pass. Initial-weight seed zero is distinct from the encryption RNG seed. FULL-2 supplies the registered accuracy gate, while FULL-1 is a separate larger-batch observation. The completed faithful encrypted subset uses $784$ inputs, the baseline's sample indices and initial weights, and its training conventions. Output width is padded to $16$, including the unused classes in the released loss; replacing this with a width-$10$ loss is not assumed equivalent. Earlier ``legacy'' runs use a different learning rate, bias policy, and initialization. Their counters and timing remain useful, but their accuracy is not comparable to the released baseline. The last subset batch is partial, affecting depth, refresh scheduling, and the effective sum-versus-mean learning-rate conversion.

The paired server measurements use the same CPU and thread cap, with different CPU parallel runtimes. Jobs are not core-pinned, and some rows overlap unrelated shared-host load. Client work and setup scopes also differ: baseline step time includes batch encryption and prediction decryption, whereas our update time excludes those separately reported client operations; both exclude key setup. Thus even a same-host ratio is not a controlled architectural speedup. We do not pool CPU and GPU samples. Encrypted rows have a single seed unless the table states otherwise; within-run iterations and Criterion samples are not independent seeds.

\subsection{RQ1: key switches and wall-clock}
\begin{table*}[t]
\centering
\scriptsize
\caption{RQ1: key switches per sample, four ways (ratio hekit / ReBoot at B = 48)}
\label{tab:rq1_ks}
\resizebox{\linewidth}{!}{%
\begin{tabular}{lrrrrrrrrrrrrr}
\toprule
Run & Input & Placement & B & Config & Samples & hekit KS/sample & Boots & ReBoot non-boot & (a) & (b) & (c) & (d) & Ways $\leq$ 0.70 \\
\midrule
\multicolumn{14}{l}{\textit{196 inputs, ReBoot full setting, 4800 samples x 10 epochs, seed 0 (headline)}} \\
FULL-1 & 196 & WN & 64 & faithful & 48000 & 18.272 & 1500 & 37 (model) & 0.232 & 0.387 & 0.395 & 0.400 & abcd \\
FULL-2 & 196 & WN & 48 in 64 & faithful & 48000 & 28.990 & 4000 & 37 (model) & 0.368 & 0.781 & 0.627 & 0.534 & acd \\
\midrule
\multicolumn{14}{l}{\textit{784 inputs, ReBoot encrypted-script subset, 1 epoch (supplementary)}} \\
F1 & 784 & WN & 32 & faithful & 1008 & 33.488 & 68 & 35.0625 (Lazy, measured) & 0.435 & 0.779 & 0.756 & 0.742 & a \\
F2 & 784 & WN & 64 & faithful & 1008 & 17.885 & 36 & 35.0625 (Lazy, measured) & 0.233 & 0.414 & 0.404 & 0.397 & abcd \\
R-41-3 & 784 & WN & 32 & legacy (R-M3-12) & 1024 & 32.531 & 64 & 35.0625 (Lazy, measured) & 0.423 & 0.741 & 0.734 & 0.730 & a \\
R-41-4 & 784 & WN & 64 & legacy (R-M3-12) & 1024 & 17.172 & 32 & 35.0625 (Lazy, measured) & 0.223 & 0.382 & 0.388 & 0.391 & abcd \\
R-41-1 & 784 & AU & 32 & legacy (R-M3-12) & 1024 & 39.469 & 128 & 35.0625 (Lazy, measured) & 0.513 & 1.149 & 0.891 & 0.730 & a \\
R-41-2 & 784 & AU & 64 & legacy (R-M3-12) & 1024 & 20.641 & 64 & 35.0625 (Lazy, measured) & 0.268 & 0.586 & 0.466 & 0.391 & abcd \\
R-41-5 & 784 & AA & 32 & legacy (R-M3-12) & 1024 & 36.000 & 96 & 35.0625 (Lazy, measured) & 0.468 & 0.945 & 0.812 & 0.730 & a \\
\midrule
\multicolumn{14}{l}{\textit{Interactive hekit, client refresh, no bootstraps (separate row, not way (d))}} \\
interactive & 784 & --- & 32 & legacy (M2) & 1024 & 19.781 & 0 & 35.0625 (Lazy, measured) & --- & --- & --- & --- & --- \\
\bottomrule
\end{tabular}}
\par\smallskip\begin{minipage}{\linewidth}\scriptsize
Independent interactive reference: ratio 0.564 against ReBoot Lazy; client refresh, not NI-derived way (d) and not a four-way pass result.\par
KS = rotations + relinearizations + SSE key switches (OpCounter). hekit bootstrap = 111 KS (measured, setup.boot\_ks). ReBoot bootstrap = 502 KS and 196-input non-boot = 37 KS/sample are M0 cost-model values (OpenFHE does not count key switches); ReBoot bootstraps 4 ciphertexts per update at B = 48 (measured on the official subset run: 84 bootstraps / 21 steps).\par
(a) own bootstrap costs; (b) both 502; (c) both 111; (d) bootstraps excluded on both sides. (b)-(d) are derived from the measured counters, not separate executions. Criterion: ratio $\leq$ 0.70 (research plan \S{}4.3.2).\par
784 rows are the supplementary subset rows (R-M3-13): ReBoot non-boot = Lazy 35.06, the in-library measurement of ReBoot's layout (Task 33, baseline c). Legacy rows use the pre-R-M3-12 config; their counts equal the faithful config but their accuracy is not comparable.\par
\end{minipage}
\end{table*}

\begin{table*}[t]
\centering
\scriptsize
\caption{Wall-clock on the same CPU (quadruped-gpu, i9-13900KF, 12 threads)}
\label{tab:wallclock_same_cpu}
\resizebox{\linewidth}{!}{%
\begin{tabular}{lrrrrrrrrr}
\toprule
Run & Input & B & Samples & Train s & s/sample & vs ReBoot & Boot s/sample & Non-boot s/sample & Threads \\
\midrule
\multicolumn{10}{l}{\textit{196 inputs, full setting (headline)}} \\
FULL-1 & 196 & 64 & 48000 & 72050.8 & 1.501 & not measured & 0.435 & 1.066 & rayon 12 \\
FULL-2 & 196 & 48 in 64 & 48000 & 146229.9 & 3.046 & not measured & 1.170 & 1.877 & rayon 12 \\
\midrule
\multicolumn{10}{l}{\textit{784 inputs, ReBoot encrypted-script subset}} \\
ReBoot official (OpenFHE) & 784 & 48 & 1008 & 3599.9 & 3.571 & 1 & 2.585 & 0.986 & OpenMP 12 \\
F1 & 784 & 32 & 1008 & 3016.7 & 2.993 & 0.838 & 0.989 & 2.003 & rayon 12 \\
F2 & 784 & 64 & 1008 & 1607.9 & 1.595 & 0.447 & 0.522 & 1.073 & rayon 12 \\
R-41-3 & 784 & 32 & 1024 & 2773.6 & 2.709 & 0.758 & 0.871 & 1.838 & rayon 12 \\
R-41-4 & 784 & 64 & 1024 & 1434.8 & 1.401 & 0.392 & 0.435 & 0.967 & rayon 12 \\
R-41-1 & 784 & 32 & 1024 & 4339.1 & 4.237 & 1.187 & 1.736 & 2.501 & rayon 12 \\
R-41-2 & 784 & 64 & 1024 & 2228.9 & 2.177 & 0.609 & 0.867 & 1.310 & rayon 12 \\
R-41-5 & 784 & 32 & 1024 & 3606.6 & 3.522 & 0.986 & 1.316 & 2.206 & rayon 12 \\
\bottomrule
\end{tabular}}
\par\smallskip\begin{minipage}{\linewidth}\scriptsize
ReBoot time = sum of its own step\_time (encrypt batch, forward/backward, decrypt prediction, weight bootstrap; setup excluded). hekit time = time\_s.train\_total (all updates including bootstraps; key generation and bootstrapper setup excluded). Present rows fail closed unless the binary's provenance names the same host and thread count as the ReBoot run.\par
Shared machine: other users' jobs (SLAM, \textasciitilde{}4 cores) ran during some rows; see docs/m3-report.md \S{}3.2. ReBoot's full 196-input setting was not run (estimated 47.6 h), so the headline has no same-CPU wall-clock ratio.\par
\end{minipage}
\end{table*}

Let $K$ be measured total key switches, $H$ bootstrap events, and $S$ examples processed. Let $c_h$ be our measured per-bootstrap cost and $c_r$ the baseline bootstrap model; let $R$ denote the baseline non-bootstrap count per example, $r$ its refreshes per update, and $B_r$ its own batch. The four reported ratios are
\begin{align}
 \rho_a&=\frac{K/S}{R+rc_r/B_r},&
 \rho_b&=\frac{(K-c_hH+c_rH)/S}{R+rc_r/B_r},\nonumber\\
 \rho_c&=\frac{K/S}{R+rc_h/B_r},&
 \rho_d&=\frac{(K-c_hH)/S}{R}.
 \label{eq:ratios}
\end{align}
Convention (a) uses each side's own bootstrap cost. Convention (b) charges both the baseline cost; (c) charges both our cost; (d) excludes bootstrap arithmetic on both sides. The interactive comparison is separate from (d): omitting bootstrap arithmetic does not remove non-interactive compact packing and unpacking. ReBoot remains at its own batch in the denominator; a hypothetical same-batch denominator would answer a different question.

The baseline's bootstrap cost remains a model because the measured OpenFHE execution does not expose key-switch counters. Its observed bootstrap frequency is used, not replaced with our schedule. The subset non-bootstrap reference is the in-library Lazy layout measurement; the full-setting reference is still a model. These boundaries are visible in Table~\ref{tab:rq1_ks}.

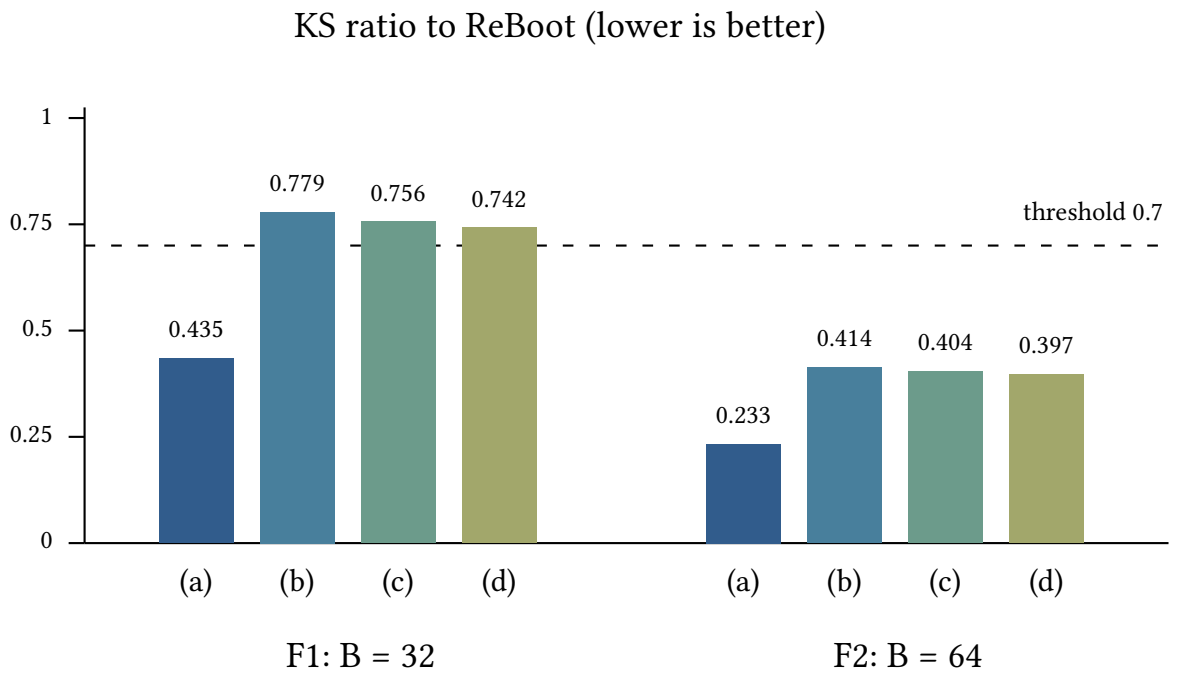
\begin{figure}[t]
\centering
\setlength{\unitlength}{1pt}
\begin{picture}(240,143)
\put(28,32){\line(1,0){204}}
\put(28,32){\line(0,1){82}}
\put(5,124){\makebox(225,10){\footnotesize KS ratio to ReBoot (lower is better)}}
\put(25,32.00){\line(1,0){3}}
\put(0,29.00){\makebox(22,7)[r]{\tiny 0}}
\put(25,52.00){\line(1,0){3}}
\put(0,49.00){\makebox(22,7)[r]{\tiny 0.25}}
\put(25,72.00){\line(1,0){3}}
\put(0,69.00){\makebox(22,7)[r]{\tiny 0.5}}
\put(25,92.00){\line(1,0){3}}
\put(0,89.00){\makebox(22,7)[r]{\tiny 0.75}}
\put(25,112.00){\line(1,0){3}}
\put(0,109.00){\makebox(22,7)[r]{\tiny 1}}
\multiput(28,88.00)(5,0){41}{\line(1,0){2}}
\put(176,91.00){\makebox(55,7)[r]{\tiny threshold 0.7}}
\put(42,32){\color[HTML]{315C8C}\rule{14pt}{34.800pt}}
\put(38,68.800){\makebox(22,7){\tiny 0.435}}
\put(42,21){\makebox(14,7){\scriptsize (a)}}
\put(61,32){\color[HTML]{487F9C}\rule{14pt}{62.320pt}}
\put(57,96.320){\makebox(22,7){\tiny 0.779}}
\put(61,21){\makebox(14,7){\scriptsize (b)}}
\put(80,32){\color[HTML]{6C9B8B}\rule{14pt}{60.480pt}}
\put(76,94.480){\makebox(22,7){\tiny 0.756}}
\put(80,21){\makebox(14,7){\scriptsize (c)}}
\put(99,32){\color[HTML]{A2A76B}\rule{14pt}{59.360pt}}
\put(95,93.360){\makebox(22,7){\tiny 0.742}}
\put(99,21){\makebox(14,7){\scriptsize (d)}}
\put(35,6){\makebox(90,10){\footnotesize F1: B = 32}}
\put(145,32){\color[HTML]{315C8C}\rule{14pt}{18.640pt}}
\put(141,52.640){\makebox(22,7){\tiny 0.233}}
\put(145,21){\makebox(14,7){\scriptsize (a)}}
\put(164,32){\color[HTML]{487F9C}\rule{14pt}{33.120pt}}
\put(160,67.120){\makebox(22,7){\tiny 0.414}}
\put(164,21){\makebox(14,7){\scriptsize (b)}}
\put(183,32){\color[HTML]{6C9B8B}\rule{14pt}{32.320pt}}
\put(179,66.320){\makebox(22,7){\tiny 0.404}}
\put(183,21){\makebox(14,7){\scriptsize (c)}}
\put(202,32){\color[HTML]{A2A76B}\rule{14pt}{31.760pt}}
\put(198,65.760){\makebox(22,7){\tiny 0.397}}
\put(202,21){\makebox(14,7){\scriptsize (d)}}
\put(138,6){\makebox(90,10){\footnotesize F2: B = 64}}
\end{picture}
\caption{Supplementary subset RQ1 ratios (one seed per row), generated from the same cells as Table~\ref{tab:rq1_ks}. Bars (a)--(d) denote the four accounting conventions in Section~\ref{sec:evaluation}. These are not equal-security or full-setting results.}
\Description{Grouped bars for the two faithful subset batches under four key-switch accounting conventions. The larger batch is below the threshold in all four; the smaller batch only under own bootstrap costs. One seed per row; not an equal-security comparison.}
\label{fig:rq1-ratios}
\end{figure}

The faithful larger-batch subset row meets the registered count threshold under every reported convention, while the smaller batch meets it only under the own-bootstrap convention. The generated figure makes that dependence visible. The corresponding larger-batch timing is also favorable on the shared CPU (Table~\ref{tab:wallclock_same_cpu}), but its subset accuracy drops relative to the baseline, including in the plaintext twin. Consequently this observation is a tradeoff, not a successful accuracy-matched RQ1 result. FULL-1 also meets the count threshold under all four conventions. FULL-2 meets conventions (a), (c), and (d), but misses the registered count target under convention (b), which charges the baseline bootstrap model on both sides. Its baseline-batch accuracy meets the registered tolerance. These successes cannot be combined across different batches into one all-conditions result. Both completed training times lack a measured full-setting ReBoot CPU denominator. The common-security condition is unmet, so the overall RQ1 goal is unmet despite the favorable conditional count comparisons.

\subsection{RQ2: placement and depth}
The AU, WN, and AA subset rows in Tables~\ref{tab:rq1_ks} and~\ref{tab:wallclock_same_cpu} provide the placement comparison. For matched legacy eMLP-1 settings, WN reduces bootstrap events and total time relative to AU. AA leaves the same lazy weight-refresh frequency and adds activation refreshes, so it does not improve this shallow workload. Its benefit is a feasible depth budget for some deeper polynomial circuits, checked separately as a structural test rather than an observed training-accuracy gain.

The corrected structural predictor equals the measured counters at every completed update. This validates accounting for the executed paths; it does not validate the baseline bootstrap model or a convergence claim. The initial cost model missed replication, key-reuse, bias, and schedule-related terms, and its optimistic interactive estimate is superseded by that explicit accounting. Both FULL-1 and FULL-2 extend exact schedule agreement to their complete registered update sequences and record accumulated weight and bias error against their respective same-batch plaintext twins. Table~\ref{tab:accuracy_vs_reboot} reports the maximum absolute final weight-or-bias difference from each twin. This diagnostic measures accumulated trajectory error, not single-bootstrap precision or an accuracy-gate threshold. That observed error is larger than in the earlier subset checks; close final accuracy does not imply a uniform small parameter error. Behavior across encrypted seeds and guarantees for arbitrary trajectories remain unestablished.

\subsection{Accuracy, local blocks, and security}
\begin{table*}[t]
\centering
\scriptsize
\caption{Test accuracy against ReBoot (R-M3-15)}
\label{tab:accuracy_vs_reboot}
\resizebox{\linewidth}{!}{%
\begin{tabular}{lrrrrrrr}
\toprule
Row & Setting & Seeds & Test acc (\%) & Twin (\%) & Reference (\%) & $\Delta$ (pt) & Gate \\
\midrule
\multicolumn{8}{l}{\textit{Full setting: 196 inputs, 4800 samples x 10 epochs, B = 48 (the $\pm$0.5 pt gate)}} \\
ReBoot released code, plaintext (recomputed twin) & full & 10 & 90.65 $\pm$ 0.27 (n = 10) & --- & --- & --- & reference \\
hekit plaintext twin (ReBoot weights, order, lr) & full & 10 & 90.65 $\pm$ 0.27 (n = 10) & --- & 90.65 $\pm$ 0.27 (n = 10) & +0.00 & twin check \\
hekit encrypted, B = 48 in 64 lanes, WN & full & 1 (seed 0) & 90.35 & 90.41 & 90.41 (seed 0) & -0.06 & PASS \\
hekit encrypted, B = 64, WN & full & 1 (seed 0) & 90.36 & 90.37 & 90.41 (seed 0) & -0.05 & not the gate (B $\neq$ 48) \\
\midrule
\multicolumn{8}{l}{\textit{eMLP-2 / eMLP-3 block MLPs, full setting, seed 0}} \\
hekit plaintext twin emlp2 (blocks 64) & full & 1 (seed 0) & 91.56 & --- & 91.56 (ReBoot twin) & +0.00 & twin check \\
hekit plaintext twin emlp3 (blocks 128,64) & full & 1 (seed 0) & 91.75 & --- & 91.75 (ReBoot twin) & +0.00 & twin check \\
\midrule
\multicolumn{8}{l}{\textit{784 inputs, ReBoot encrypted-script subset, 1008 samples, 1 epoch (not a gate, R-M3-15)}} \\
ReBoot official encrypted (OpenFHE) & subset, B = 48 & 1 (seed 0) & 75.06 & --- & --- & --- & reference \\
hekit encrypted WN, B = 32 & subset & 1 (seed 0) & 75.77 & 75.77 & 75.06 & +0.71 & not a gate \\
hekit encrypted WN, B = 64 & subset & 1 (seed 0) & 73.58 & 73.59 & 75.06 & -1.48 & not a gate \\
hekit encrypted eMLP-2, B = 48 (196) & subset 1008 & 1 & pending (R-52-1) & pending (R-52-1) &  &  & twin check \\
hekit encrypted eMLP-3, B = 48 (196) & subset 1008 & 1 & pending (R-52-2) & pending (R-52-2) &  &  & twin check \\
\midrule
\multicolumn{8}{l}{\textit{Reference only: the paper value and the released code at the paper setting (R-REB-3)}} \\
ReBoot paper, Table 4 (plaintext) & 60k samples & 10 (paper) & 94.61 $\pm$ 0.27 & --- & --- & --- & not reproducible from the released defaults \\
ReBoot released code, paper setting, constant lr, epoch 50 & 60k samples, 50 epochs & 10 & 95.15 $\pm$ 0.19 (n = 10) & --- & 94.61 & +0.54 & reference \\
ReBoot released code, paper setting, constant lr, best epoch (test-selected) & 60k samples, 50 epochs & 10 & 95.35 $\pm$ 0.10 (n = 10) & --- & 94.61 & +0.74 & reference \\
ReBoot released code, paper setting, ReBoot's scheduler, epoch 50 & 60k samples, 50 epochs & 10 & 95.25 $\pm$ 0.08 (n = 10) & --- & 94.61 & +0.64 & reference \\
ReBoot released code, paper setting, ReBoot's scheduler, best epoch (test-selected) & 60k samples, 50 epochs & 10 & 95.46 $\pm$ 0.09 (n = 10) & --- & 94.61 & +0.85 & reference \\
\bottomrule
\end{tabular}}
\par\smallskip\begin{minipage}{\linewidth}\scriptsize
FULL-2: maximum final weight/bias error vs same-batch plaintext twin = 0.00662503; trajectory diagnostic, not per-bootstrap precision or an accuracy gate.\par
FULL-1: maximum final weight/bias error vs same-batch plaintext twin = 0.00682657; trajectory diagnostic, not per-bootstrap precision or an accuracy gate.\par
R-M3-15: the $\pm$0.5 pt gate compares hekit encrypted B = 48 (FULL-2) with ReBoot's released code at its full setting, seed 0. Subset rows are 1 epoch and are not a gate. The paper value and R-REB-3 are reported with this note only.\par
\end{minipage}
\end{table*}

\begin{table*}[t]
\centering
\scriptsize
\caption{Plaintext dataset reproduction against ReBoot Table 4 (Task 51)}
\label{tab:plain_datasets}
\resizebox{\linewidth}{!}{%
\begin{tabular}{lrrrrrrrr}
\toprule
Dataset & Config & Epochs & Seeds & Fixed final epoch (\%) & Best epoch, test-selected (\%) & ReBoot Table 4 (\%) & $\Delta$ fixed (pt) & Within 1 pt \\
\midrule
MNIST 14x14 & paper setting & 50 & 10 & 95.00 $\pm$ 0.18 (n = 10) & 95.33 $\pm$ 0.15 (n = 10) & 94.61 $\pm$ 0.27 & +0.39 & yes \\
Fashion-MNIST 14x14 & paper setting & 50 & 10 & 85.22 $\pm$ 0.17 (n = 10) & 85.53 $\pm$ 0.12 (n = 10) & 85.12 $\pm$ 0.20 & +0.10 & yes \\
Breast Cancer & fixed split, min-max (primary) & 100 & 10 & 96.40 $\pm$ 1.27 (n = 10) & 99.30 $\pm$ 0.37 (n = 10) & 99.03 $\pm$ 1.78 & -2.63 & no \\
Breast Cancer & per-run split (sensitivity) & 100 & 10 & 95.26 $\pm$ 1.61 (n = 10) & 97.89 $\pm$ 0.94 (n = 10) & 99.03 $\pm$ 1.78 & -3.77 & no \\
Breast Cancer & z-score (diverges) & 100 & 10 & diverged (10/10 seeds) & --- & 99.03 $\pm$ 1.78 & --- & no \\
\bottomrule
\end{tabular}}
\par\smallskip\begin{minipage}{\linewidth}\scriptsize
Fixed final epoch is the primary figure. The best epoch is selected on the test set and is optimistic; it is shown only because ReBoot does not state its epoch count. Seeds: the plain binary's own seeds (not ReBoot's initial weights).\par
Fashion-MNIST and Breast Cancer are plaintext twins only (no encrypted runs).\par
\end{minipage}
\end{table*}

\begin{table*}[t]
\centering
\scriptsize
\caption{Security estimates (lattice-estimator; log2 rop, minimum over attacks)}
\label{tab:security}
\resizebox{\linewidth}{!}{%
\begin{tabular}{lrrrrrr}
\toprule
Parameters & N & log PQ & Secret & BDGL16 & MATZOV & >= 128 (both) \\
\midrule
hekit (a): n16\_boot main key & 2\textasciicircum{}16 & 1732 (1731.998) & dense ternary (P($\pm$1) = 1/4 each, h $\approx$ N/2) & 131.5 & 129.3 & yes \\
hekit (b): SSE key s -> s' (h' = 32) & 2\textasciicircum{}16 & 294 (q\_0$\cdot$P) & sparse ternary, h' = 32 & 133.8 & 134.2 & yes \\
ReBoot main key (R-REB-4: OpenFHE's own chain) & 2\textasciicircum{}16 & 1730 (1729.585) & sparse ternary, h = 192 (97 / 95) & 121.8 & 121.5 & no \\
\bottomrule
\end{tabular}}
\par\smallskip\begin{minipage}{\linewidth}\scriptsize
ReBoot row: the MITM tier (operation counts; about 2\textasciicircum{}104-2\textasciicircum{}110 memory). Without MITM attacks ReBoot is at 127.8 / 126.2 (BDGL16 / MATZOV), also below 128.\par
Source: the summary table of docs/security.md (the estimator outputs are recorded in that doc; no JSON exists). Same lattice-estimator models for every row; m = N; Core-SVP is in the doc for information only. Equal N does not mean equal security: no equal-security claim.\par
\end{minipage}
\end{table*}

Table~\ref{tab:accuracy_vs_reboot} separates the released full-setting reference, subset results, and published paper accuracy. The released defaults and paper-setting reconstruction are different protocols. We do not choose whichever reference makes the result look best. FULL-2 meets the registered $\pm0.5$ percentage-point tolerance against the captured released full-setting seed-zero reference. The comparison uses that individual reference, not the released ten-seed mean or the paper-setting accuracy. FULL-1 reports its larger-batch accuracy and same-batch twin separately, without substituting either for this gate. Matching a plaintext twin is necessary to diagnose arithmetic fidelity, but is insufficient to establish baseline-equivalent generalization.

Plaintext dataset experiments (Table~\ref{tab:plain_datasets}) provide broader optimization checks. MNIST and Fashion-MNIST reproduce the stated comparison tolerance at the fixed final epoch. Breast Cancer does not: the fixed-epoch min-max result misses its target, and selecting the best test epoch is optimistic leakage from model selection. The z-score setting diverges for the polynomial activation. Fashion-MNIST and Breast Cancer results are plaintext twins, not encrypted training measurements. The local-block plaintext twins reproduce the released eMLP-2/eMLP-3 values for the recorded seed; their encrypted subset rows are pending. The block experiments do not extend the current headline claim to deeper networks.

Table~\ref{tab:security} applies common estimator models to the actual exposed key moduli and secret distributions. The evaluated main and SSE instances of our preset exceed the chosen operation-count target under those models; the released ReBoot preset does not. This is an estimate over the documented attack search, not a security proof. The smallest sparse-secret values use attacks with extreme memory requirements, and even non-MITM minima have substantial or unreported memory demands. These are not practical attack demonstrations. Equal $N$ and a requested library security label do not establish equal security. Results must remain conditional until either a security-different comparison is explicitly accepted or an independently estimated and remeasured alternative baseline is available.

\subsection{RQ3: device operations and ablation}
\begin{table*}[t]
\centering
\scriptsize
\caption{GPU vs CPU vs HEonGPU, n16\_train (historical R-47-2 Criterion; R-47-3 example medians, ms)}
\label{tab:gpu_vs_cpu_ops}
\resizebox{\linewidth}{!}{%
\begin{tabular}{lrrrrrrr}
\toprule
Operation & Level & hekit GPU & CPU 1 thr & CPU 12 thr & GPU speed-up (1 / 12 thr) & HEonGPU (RTX 4090) & GPU / HEonGPU \\
\midrule
add & 31 & 0.111 & 15.674 & 14.611 & 141x / 132x & none published & --- \\
relinearize & 31 & 4.671 & 639.639 & 240.070 & 137x / 51x & 2.582 (l = 36, I = 4 primes per digit) & 1.81x (nominal) \\
mul + relin + rescale & 31 & 5.823 & 725.209 & 337.731 & 125x / 58x & none published & --- \\
rotate(1) & 31 & 3.990 & 622.046 & 223.880 & 156x / 56x & none published & --- \\
hoisted rotate x16 & 31 & skipped: needs 5580 MiB (2377 MiB free) & 3765.068 & 689.471 & --- & none published & --- \\
hoisted rotate x16 (capped keys) & 9 & 4.984 & 1005.331 & 172.870 & 202x / 35x & none published & --- \\
matvec\_enc d = 64 & 31 & skipped: needs 7276 MiB (1929 MiB free) & 12938.440 & 2325.173 & --- & none published & --- \\
matvec\_enc d = 64 (capped keys) & 9 & 16.120 & 2523.223 & 547.762 & 157x / 34x & none published & --- \\
matvec\_enc d = 64 (capped keys) & 5 & 9.509 & 1391.064 & 314.649 & 146x / 33x & none published & --- \\
matvec\_enc x4, 1 stream (sequential) & 5 & 34.994 & --- & --- & --- & none published & --- \\
matvec\_enc x4, 4 streams (untuned R-47-2) & 5 & 41.673 & --- & --- & --- & none published & --- \\
matvec\_enc sequential x4, 1 stream (R-47-3, pool default) & 5 & 36.773 & --- & --- & --- & none published & --- \\
matvec\_enc batch x4, 2 streams (R-47-3, pool default) & 5 & 32.080 & --- & --- & --- & none published & --- \\
matvec\_enc batch x4, 4 streams (R-47-3, pool default) & 5 & 32.889 & --- & --- & --- & none published & --- \\
matvec\_enc sequential x4, 1 stream (R-47-3, pool max) & 5 & 34.784 & --- & --- & --- & none published & --- \\
matvec\_enc batch x4, 2 streams (R-47-3, pool max) & 5 & 29.173 & --- & --- & --- & none published & --- \\
matvec\_enc batch x4, 4 streams (R-47-3, pool max) & 5 & 28.821 & --- & --- & --- & none published & --- \\
\bottomrule
\end{tabular}}
\par\smallskip\begin{minipage}{\linewidth}\scriptsize
matvec\_enc sequential x4, 1 stream (R-47-3, pool default): min 36.693 ms; SD 0.203 ms; 30 within-run repetitions.\par
matvec\_enc batch x4, 2 streams (R-47-3, pool default): min 31.841 ms; SD 0.652 ms; 30 within-run repetitions.\par
matvec\_enc batch x4, 4 streams (R-47-3, pool default): min 32.087 ms; SD 0.235 ms; 30 within-run repetitions.\par
matvec\_enc sequential x4, 1 stream (R-47-3, pool max): min 34.664 ms; SD 0.209 ms; 30 within-run repetitions.\par
matvec\_enc batch x4, 2 streams (R-47-3, pool max): min 28.940 ms; SD 0.222 ms; 30 within-run repetitions.\par
matvec\_enc batch x4, 4 streams (R-47-3, pool max): min 28.502 ms; SD 0.227 ms; 30 within-run repetitions.\par
R-47-3 uses example medians, minima and population SD over 30 timed repetitions after 3 warmups, with device synchronization inside each timing; not 30 independent seeds. Successful child completion includes the producer assertions of batch/sequential residue equality. CPU setup seconds are not CPU matvec latency; these are not end-to-end training and not GPU peak measurements. R-47-2 Criterion rows remain historical; no unlike-run CPU speedups or HEonGPU ratios are computed for R-47-3.\par
R-47-3 original terminal provenance is checked here; scientific acceptance remains a separate review. Input capture-time SHA256 (not historical execution proof): results/quadruped/raw/R-47-3/R-47-3-example/program.log = 7ce4fe8d3a63f3391117ae050e1cfbf200d0131d5902315cb59ddf3fdd54e3a5; results/quadruped/raw/R-47-3/R-47-3-example/meta.json = af432f72f440560911e16feebfc047258c822f756fdb6cc234c713eaad100a49; results/quadruped/raw/R-47-3/R-47-3-example/post.txt = 87927caf5dd61c87a02f350f687e37f490a709e4dd5d090aae28bd2e3fb76a9d; results/quadruped/raw/R-47-3/R-47-3-example/cmd.txt = f60c0ae8a42744499de58ff24c95db487dd8af0afc63c49feca3b5313d68f54c; results/quadruped/raw/R-47-3/R-47-3-example/state.json = 4e6a087ccd329bf1e25585c4f44fa36abcdc484f42eadc92a01597053e5b475a\par
HEonGPU relinearize: Ozcan and Savas, GPU Implementations of Three Different Key-Switching Methods for Homomorphic Encryption Schemes, IACR ePrint 2025/124, Table 3: CKKS relinearization on a GeForce RTX 4090, log N = 16, 43-bit precision, Method II (hybrid), (I, l\textasciitilde{}, l) = (4, 40, 36), level 35. Author-deposited manuscript dated November 2024, verified in Sabanci University repository 53894 on 2026-11-08. Caveat: not the same parameters as n16\_train (36 vs 32 Q primes, 4 special primes; I = 4 primes per digit corresponds to alpha = 4, not four digits; n16\_train uses dnum 8); the ratio is nominal. Scaling key switching linearly in l gives about 2.30 ms at l = 32 and a ratio of about 2.0x (docs/m4-report.md section 1.1). ePrint 2024/1543 could not be read (HTTP 403), so other HEonGPU operation values are unconfirmed. The source states that Method II/III use lower security than its 128-bit Method I comparison, with an approximately 110-bit minimum across experiments; this is not a matched-security comparison.\par
Capped keys: the input is dropped to level 9 or 5 and the keys are cut to that level (faithful for that level, not the top-level cost); the CPU uses full keys at the same level. The shared RTX 4090 had 2-3 GiB free, so full-key GPU rows that need more were skipped.\par
\end{minipage}
\end{table*}

\begin{table*}[t]
\centering
\scriptsize
\caption{RQ3 ablation: one interactive eMLP-1 iteration under each EvalConfig}
\label{tab:rq3_ablation}
\resizebox{\linewidth}{!}{%
\begin{tabular}{lrrrrrrrrr}
\toprule
Source & Backend / level & hoist bsgs lazy & Rotations & Relins & KS/iter & Closed form & Step median (s) & Step / default & Max |$\Delta$W| vs default \\
\midrule
CPU, n16\_train, level 31 & cpu, level max & 1 1 1 & 571 & 62 & 633 & match & 134.040 & 1.000 & 0 \\
CPU, n16\_train, level 31 & cpu, level max & 1 1 0 & 571 & 114 & 685 & match & 137.170 & 1.023 & 4.81e-07 \\
CPU, n16\_train, level 31 & cpu, level max & 1 0 1 & 610 & 101 & 711 & match & 142.124 & 1.060 & 4.14e-07 \\
CPU, n16\_train, level 31 & cpu, level max & 1 0 0 & 610 & 114 & 724 & match & 143.025 & 1.067 & 3.60e-07 \\
CPU, n16\_train, level 31 & cpu, level max & 0 1 1 & 571 & 62 & 633 & match & 134.069 & 1.000 & 0 \\
CPU, n16\_train, level 31 & cpu, level max & 0 1 0 & 571 & 114 & 685 & match & 132.606 & 0.989 & 4.81e-07 \\
CPU, n16\_train, level 31 & cpu, level max & 0 0 1 & 610 & 101 & 711 & match & 136.063 & 1.015 & 4.14e-07 \\
CPU, n16\_train, level 31 & cpu, level max & 0 0 0 & 610 & 114 & 724 & match & 136.274 & 1.017 & 3.60e-07 \\
GPU, n16\_train, level 7 & NVIDIA GeForce RTX 4090 (sm\_89), level 7 & 1 1 1 & 571 & 62 & 633 & match & 0.172 & 1.000 & 0 \\
GPU, n16\_train, level 7 & NVIDIA GeForce RTX 4090 (sm\_89), level 7 & 1 1 0 & 571 & 114 & 685 & match & 0.185 & 1.075 & 5.32e-07 \\
GPU, n16\_train, level 7 & NVIDIA GeForce RTX 4090 (sm\_89), level 7 & 1 0 1 & 610 & 101 & 711 & match & 0.196 & 1.135 & 5.44e-07 \\
GPU, n16\_train, level 7 & NVIDIA GeForce RTX 4090 (sm\_89), level 7 & 1 0 0 & 610 & 114 & 724 & match & 0.199 & 1.156 & 5.18e-07 \\
GPU, n16\_train, level 7 & NVIDIA GeForce RTX 4090 (sm\_89), level 7 & 0 1 1 & 571 & 62 & 633 & match & 0.169 & 0.980 & 0 \\
GPU, n16\_train, level 7 & NVIDIA GeForce RTX 4090 (sm\_89), level 7 & 0 1 0 & 571 & 114 & 685 & match & 0.188 & 1.090 & 5.32e-07 \\
GPU, n16\_train, level 7 & NVIDIA GeForce RTX 4090 (sm\_89), level 7 & 0 0 1 & 610 & 101 & 711 & match & 0.204 & 1.182 & 5.44e-07 \\
GPU, n16\_train, level 7 & NVIDIA GeForce RTX 4090 (sm\_89), level 7 & 0 0 0 & 610 & 114 & 724 & match & 0.202 & 1.174 & 5.18e-07 \\
CPU, n16\_train, level 7 & cpu, level 7 & 1 1 1 & 571 & 62 & 633 & match & 16.822 & 1.000 & 0 \\
CPU, n16\_train, level 7 & cpu, level 7 & 1 1 0 & 571 & 114 & 685 & match & 17.788 & 1.057 & 5.32e-07 \\
CPU, n16\_train, level 7 & cpu, level 7 & 1 0 1 & 610 & 101 & 711 & match & 18.914 & 1.124 & 5.44e-07 \\
CPU, n16\_train, level 7 & cpu, level 7 & 1 0 0 & 610 & 114 & 724 & match & 18.751 & 1.115 & 5.18e-07 \\
CPU, n16\_train, level 7 & cpu, level 7 & 0 1 1 & 571 & 62 & 633 & match & 17.168 & 1.021 & 0 \\
CPU, n16\_train, level 7 & cpu, level 7 & 0 1 0 & 571 & 114 & 685 & match & 18.035 & 1.072 & 5.32e-07 \\
CPU, n16\_train, level 7 & cpu, level 7 & 0 0 1 & 610 & 101 & 711 & match & 18.859 & 1.121 & 5.44e-07 \\
CPU, n16\_train, level 7 & cpu, level 7 & 0 0 0 & 610 & 114 & 724 & match & 18.716 & 1.113 & 5.18e-07 \\
CPU smoke (insecure N = 2\textasciicircum{}12: counts only) & cpu, level max & 1 1 1 & 171 & 62 & 233 & match & n/a (smoke) & n/a (smoke) & 0 \\
CPU smoke (insecure N = 2\textasciicircum{}12: counts only) & cpu, level max & 1 1 0 & 171 & 114 & 285 & match & n/a (smoke) & n/a (smoke) & 7.83e-09 \\
CPU smoke (insecure N = 2\textasciicircum{}12: counts only) & cpu, level max & 1 0 1 & 210 & 101 & 311 & match & n/a (smoke) & n/a (smoke) & 6.90e-09 \\
CPU smoke (insecure N = 2\textasciicircum{}12: counts only) & cpu, level max & 1 0 0 & 210 & 114 & 324 & match & n/a (smoke) & n/a (smoke) & 7.39e-09 \\
CPU smoke (insecure N = 2\textasciicircum{}12: counts only) & cpu, level max & 0 1 1 & 171 & 62 & 233 & match & n/a (smoke) & n/a (smoke) & 0 \\
CPU smoke (insecure N = 2\textasciicircum{}12: counts only) & cpu, level max & 0 1 0 & 171 & 114 & 285 & match & n/a (smoke) & n/a (smoke) & 7.83e-09 \\
CPU smoke (insecure N = 2\textasciicircum{}12: counts only) & cpu, level max & 0 0 1 & 210 & 101 & 311 & match & n/a (smoke) & n/a (smoke) & 6.90e-09 \\
CPU smoke (insecure N = 2\textasciicircum{}12: counts only) & cpu, level max & 0 0 0 & 210 & 114 & 324 & match & n/a (smoke) & n/a (smoke) & 7.39e-09 \\
\bottomrule
\end{tabular}}
\par\smallskip\begin{minipage}{\linewidth}\scriptsize
R-48-4: success; optimizer nag; child exit 0; watchdog false. Runner/queue fields and hashes identify captured bytes, not historical execution proof.\par
R-48-4: one warmup; 3 timed iterations within seed 48, not independent seeds. Step is server work including device-queue wait; iteration\_s\_median is the sum of step and refresh medians, not a median of paired iteration totals. These are interactive client-refresh measurements, not NI bootstrap and not GPU peak measurements.\par
R-48-4: rayon (feature parallel), RAYON\_NUM\_THREADS = 12, 32 logical CPUs; started 1794139512.8188622; finished 1794145084.3839154; load\_start 2.26 2.49 2.46 3/1430 2393700; load\_end 2.26 2.88 3.01 2/1423 2409266; original binary SHA256 5044b5fe9c31f837f0ccff325267756dcfeae77931c6df059d49de44cf52eb67.\par
CPU, n16\_train, level 31: optimizer nag, B = 32, layout d = 1024, start level 31. Rows with unlike conditions are not a CPU/GPU speed comparison.\par
R-48-2: success; optimizer nag; child exit 0; watchdog false. Runner/queue fields and hashes identify captured bytes, not historical execution proof.\par
R-48-2: one warmup; 5 timed iterations within seed 48, not independent seeds. Step is server work including device-queue wait; iteration\_s\_median is the sum of step and refresh medians, not a median of paired iteration totals. These are interactive client-refresh measurements, not NI bootstrap and not GPU peak measurements.\par
R-48-2: rayon (feature parallel), RAYON\_NUM\_THREADS = 12, 32 logical CPUs; started 1794137697.7851274; finished 1794138183.439248; load\_start 2.05 2.56 2.60 3/1432 2388098; load\_end 2.13 2.20 2.40 2/1433 2389482; original binary SHA256 95db0da443234d5d6bcb04b4891e50df526ff1944008edac4e452c86c54d2aab.\par
GPU, n16\_train, level 7: optimizer nag, B = 32, layout d = 1024, start level 7. Rows with unlike conditions are not a CPU/GPU speed comparison.\par
R-48-3: success; optimizer nag; child exit 0; watchdog false. Runner/queue fields and hashes identify captured bytes, not historical execution proof.\par
R-48-3: one warmup; 5 timed iterations within seed 48, not independent seeds. Step is server work including device-queue wait; iteration\_s\_median is the sum of step and refresh medians, not a median of paired iteration totals. These are interactive client-refresh measurements, not NI bootstrap and not GPU peak measurements.\par
R-48-3: rayon (feature parallel), RAYON\_NUM\_THREADS = 12, 32 logical CPUs; started 1794138192.776057; finished 1794139506.0015738; load\_start 2.03 2.18 2.39 3/1431 2389520; load\_end 2.37 2.51 2.47 2/1442 2393664; original binary SHA256 cccf1c7b397be7176abfd89790b5c0bfbb47b5b2567e2b1cf2db94d66a9b1299.\par
CPU, n16\_train, level 7: optimizer nag, B = 32, layout d = 1024, start level 7. Rows with unlike conditions are not a CPU/GPU speed comparison.\par
CPU smoke: optimizer nag, B = 2, layout d = 1024, start level 10. Rows with unlike conditions are not a CPU/GPU speed comparison.\par
1. Definition of bsgs = false. The giant rotation is not shared; the weights keep the same pre-rotated (BSGS) layout and the same Galois keys. Every product with a nonzero giant shift is finished (relinearized) and rotated on its own; the products of the zero-shift block are summed as with BSGS. Per matvec of dimension d (offset 0): d - 1 rotations, of which only the baby - 1 rotations of x can be hoisted; with lazy\_relin 1 + d - baby relins, without it d (EvalConfig::matvec\_rotations, EvalConfig::matvec\_enc\_relins). At d = 16 (baby = giant = 4), as (hoistable rotations / individual rotations / relins): (form / hoistable rot. / individual rot. / relins (lazy) / relins (not lazy));    (default BSGS / 3 / 3 / 4 / 16);  (bsgs = false (this library) / 3 / 12 / 13 / 16);  (textbook diagonal method (not implemented) / 15 / 0 / 1 / 16);  The textbook (Halevi-Shoup) diagonal method rotates x by all d - 1 steps (all hoistable) and sums the products lazily; it needs un-pre-rotated diagonals, i.e. another weight layout and gradient form, so it is not implemented. The all-off configuration has the same counts in both forms (15 rotations, 16 relins at d = 16).\par
2. Bias. Against the strongest naive baseline (the textbook method with hoisting and lazy relinearization), the cost of the bsgs = false rows overstates what BSGS gains: the measured BSGS gain is an upper bound. Conversely, the marginal gains of hoist and lazy\_relin measured inside bsgs = false configurations are lower bounds (only baby - 1 rotations can be hoisted and only the zero-shift block summed lazily: e.g. lazy\_relin saves 3 relins at d = 16 instead of 15).\par
3. hoist changes no OpCounter value. The counter does not distinguish hoisted from individual rotations, so the effect of hoist shows only in wall-clock time, never in key-switch counts.\par
4. bsgs does not touch the gradient's outer product (Dense::weight\_grad\_scaled, Dense::weight\_grad\_reduced): its (baby - 1) + (giant - 1) input rotations are what the pre-rotated weight layout requires, not a BSGS choice; only hoist changes how they run. The interactive gradient multiplies each diagonal with one Evaluator::mul (nothing to sum lazily), so lazy\_relin does not change it either.\par
The smoke rows (insecure N = 2\textasciicircum{}12, high Mac load) check counts and closed forms only; their times are not comparative.\par
\end{minipage}
\end{table*}

Table~\ref{tab:gpu_vs_cpu_ops} contains synchronized operation latencies, including capped-level rows and memory skips. The HEonGPU relinearization reference~\cite{gpuKeys2025} is nominal: its modulus count differs. A linear-in-limb adjustment is a cost-model extrapolation, not an equal-parameter measurement. We verified the cited relinearization row against the author-deposited manuscript. Its Method II/III configurations use lower security than its Method I comparison, so this reference is not a matched-security speed comparison. The repaired pool/submission path is measured separately by R-47-3: four independent level-$5$ products, three warmups and thirty synchronized repetitions per condition. Each pool threshold shows a latency reduction from batching relative to its own sequential row. These are within-run descriptive comparisons, not independent seeds or isolated estimates of the individual code changes. The profiled five-repetition run is excluded from the timing table because instrumentation changes the measurement. Its trace remains unanalyzed. The accompanying test process reports $45$ successful test functions, including three memory-capacity skips; only $42$ executed, including expected-error paths. Full-key skipped cases and GPU peak memory remain unmeasured.

Table~\ref{tab:rq3_ablation} includes the recovered R-48 server measurements. The matched level-$7$ CPU/CUDA rows use a $784\to32\to10$ network, NAG, and batch $32$, with one warmup and five timed iterations within seed $48$. Default server-step medians are $16.8223$~s on CPU and $0.172313$~s on RTX~4090, a descriptive $97.63\times$ ratio excluding interactive client refresh. The level-$31$ CPU rows are not a matched GPU comparison. The insecure small-parameter CPU smoke validates counts and checks only; its high-load timings are inadmissible for performance comparisons. The flag named \texttt{bsgs=false} keeps pre-rotated weights rather than implementing the strongest textbook diagonal baseline. Thus its BSGS gain would be an upper bound against that alternative, while the marginal hoisting and lazy-relinearization gains within those rows would be lower bounds. Hoisting changes no operation counter; the gradient outer product retains layout-required rotations independent of the BSGS flag.

A separate retained level-$7$ CPU/CUDA training pair uses the same $784\to32\to10$ architecture, batch $32$, and initial-weight seed $32$ for $32$ updates. All $25{,}450$ final weight and bias values are numerically equal between the retained outputs, as are the per-batch operation counters. Both models obtain $29\%$ accuracy on $1{,}000$ test examples after this short run; this is not converged accuracy. Mean server-step times are $16.8016$~s and $0.167773$~s, respectively; including client refresh and transfers gives $24.4458$~s and $8.95422$~s ($2.73\times$). These are interactive measurements without bootstrapping. Runner records and captured-byte hashes identify the retained evidence, but the historical training JSON lacks embedded executable/input provenance. This bounded agreement does not establish full-training accuracy preservation or the speed of the main $196\to32\to16$ non-interactive workload.

\paragraph{Matched non-interactive training with device bootstrap.}
\begin{table}[t]
\centering\small
\caption{Matched one-epoch training with device-resident CUDA bootstrap.}
\label{tab:gpu_epoch}
\begin{tabular}{lrr}
\toprule
Metric & CPU & CUDA \\
\midrule
Mean update incl. refresh (s) & 142.438 & 2.670 \\
Mean refresh work/update (s) & 55.430 & 1.859 \\
One-epoch training (s) & 14268.295 & 293.075 \\
Total incl. setup/evaluation (s) & 14325.222 & 352.247 \\
Test accuracy (\%) & 81.92 & 81.92 \\
\bottomrule
\end{tabular}
\par\smallskip\begin{minipage}{\linewidth}\footnotesize
$196\to32\to16$, $B=48$, 4,800 ordered training images, 100 updates and 400 bootstraps per backend. Same executable, inputs and 12-thread cap; sequential runs on the same host. Test accuracy uses decrypted trained weights in plaintext on 10,000 examples. This is one epoch of the selected 4,800-image workload, not the entire 60,000-image training set or a ten-epoch accuracy result.
\end{minipage}
\end{table}

The main $196\to32\to16$ network uses $B=48$, initial-weight seed zero, and the FULL-2 sample order. Table~\ref{tab:gpu_epoch} reports a matched 100-update CPU/CUDA pair using the same executable, input hashes, cryptographic parameters and $12$-thread cap, sequentially on the Core i9-13900KF/RTX~4090 host. Each update performs four non-interactive bootstraps, with ciphertext arithmetic on the GPU in the CUDA path. Training totals are $14{,}268.295$~s and $293.075$~s, a $48.68\times$ ratio against our CPU implementation. Mean synchronized updates yield $53.35\times$; setup- and evaluation-inclusive totals yield $40.67\times$. Both runs execute $400$ bootstraps and $136{,}900$ key switches, with every predicted update count matched. This is one pass over the selected $4{,}800$ images, not all $60{,}000$ MNIST training examples. Within-run updates are not independent seed replications.

All $6{,}832$ decoded parameters and $160{,}000$ test-output values agree exactly. Both obtain $81.92\%$ accuracy on the same $10{,}000$ test images, versus $81.91\%$ for the plaintext twin; maximum parameter error against the twin is $6.5269\times10^{-4}$. Accuracy and MSE are independently recomputed from retained outputs and pinned labels. Test evaluation uses decrypted trained weights in plaintext. The earlier same-executable two-update gate yielded $49.36\times$ training acceleration and exact decoded parity. A historical executable using CUDA arithmetic with CPU bootstrap required $56.626$~s per update, including $55.808$~s of refresh and adapter work. It motivated the device path but is not a same-executable three-way ablation.

\paragraph{Completed ten-pass GPU training.}
After the matched pair, the same GPU executable completed the original $1{,}000$-update workload with the original variable learning-rate schedule: ten passes over $4{,}800$ images, $4{,}000$ bootstraps, and $1{,}391{,}540$ key switches. Every per-update counter prediction matched. Actual training took $2{,}967.095$~s (49.45 minutes), with $3{,}026.138$~s including setup and final evaluation. The fixed final model achieves $90.35\%$ test accuracy versus $90.41\%$ for its plaintext twin; maximum parameter error is $6.6250\times10^{-3}$. This is an actual completed run, not a projection from the first pass. Its accuracy equals historical CPU FULL-2 and meets the registered tolerance against the released seed-zero reference. However, historical FULL-2 uses a different executable and exports no final weights. It therefore supplies neither exact full-run parameter parity nor a matched full-run speed ratio. The 100-update pair supplies the controlled timing comparison; the full GPU run supplies the longer-horizon accuracy observation, each at one initialization.

CPU and GPU training comparisons require the same optimizer, level, layout, and batch. Registered SGD fallback rows are compared only with the corresponding CPU SGD row. A resource skip is unmeasured; a device OOM is a failure. Test-process success cannot stand in for an actually executed production-size parity gate. The executed full-slot GPU bootstrap gate measures precision and one refresh latency (Section~\ref{sec:gpu}); the $220$~ms target is missed. Pool checkpoints do not establish peak memory, and a single refresh does not characterize a latency distribution. HEonGPU's separate bootstrap reference~\cite{heongpuBench} has a different remaining-level budget and main-secret distribution; the fixed plan target is not a normalized speed-ratio result. Repeated independent runs and a matched full-run CPU/GPU parameter comparison remain future work.

\section{Related Work, Limitations, and Conclusion}
\label{sec:limits}
\subsection{Positioning}
The earlier diagonal-training design~\cite{mihara2020} motivates rotation-only transposition. We generalize the layout and charge server-side batch maintenance, compact refresh, and optimizer state. ReBoot~\cite{reboot2026} already establishes non-interactive training and local learning; its released implementation is our application baseline, not a source of imported implementation code. HETAL~\cite{hetal2023} studies encrypted transfer learning and efficient matrix products in a different training scope. Our evaluation does not reproduce its workloads or establish a ranking against it.

The cryptographic building blocks are efficient linear transforms and hoisting~\cite{halevi2018}, SSE~\cite{sse2022}, and CKKS~\cite{ckks2017}. SEAL~\cite{seal} provides interoperability checks; OpenFHE~\cite{openfhe2022} supplies the separately executed baseline. HEonGPU~\cite{gpuKeys2025,heongpuBench} is a device-operation reference, not an end-to-end training competitor measured under our protocol. An MIT implementation improves reuse, but licensing and pipeline completeness are not evidence of a cryptographic or learning-performance advance. The approved dependency policy permits MIT, Apache, BSD, and ISC licenses, with one narrow Unicode-3.0 character-table exception for \texttt{unicode-ident} version \texttt{1.0.27}. That component requires its complete Unicode notice; this is not a general Unicode-license exception.

\subsection{Limits and missed goals}
\paragraph{Scope and comparison.}
FULL-1 and FULL-2 supply completed full-setting encrypted runs at two batches with the same initial-weight seed, including their schedules and accumulated error. FULL-2 passes the registered baseline-batch accuracy gate, but misses the common-baseline-bootstrap count target. FULL-1's successful count comparisons are a separate configuration. Full-setting ReBoot CPU wall-clock is unmeasured, and the common-security condition is unmet. Neither these two configurations nor the earlier subsets establish multi-seed encrypted behavior; multi-seed plaintext results are a different evidence class. The observed parameter error does not establish a general error bound. Batch choices trade refresh amortization against learning dynamics. The effective baseline security differs, initialization and bias arithmetic require care, and legacy rows cannot support accuracy claims. ReBoot's in-library layout timing was measured on a different machine and is not a same-CPU baseline; its public-label/client-update path has a weaker confidentiality contract. The official encrypted block-model baselines have not been run: their reported memory demand exceeds the available host budget.

\paragraph{Cost and correctness.}
A key switch at a low level need not cost the same as one at the top level; counts omit memory traffic, cache effects, and key residency. The initial compact-layout assumptions required actual CKKS tests and the first model was optimistic. Replicated copies were once unstable during interactive training; canonical-layout refresh was required to avoid divergence. The earlier paper's published timing total and phase sum disagree, so the continuity demonstration does not silently reconcile them or infer a server-only speedup. The original CPU NTT target and several SEAL-relative key-switch performance targets were missed. The early CUDA NTT spike was not timed against its registered hardware reference; later device-operator measurements do not retroactively supply that measurement. Some SEAL comparisons also have different special-prime/key-switch decompositions. We report these as missed performance targets, not failed numerical gates.

\paragraph{Numerical limits.}
SSE and modulus-band changes were needed because the planned dense-secret overflow range and original prime allocation did not meet the bootstrap precision budget. Unit-scale nonconstant masks were invalid, and exact-scale folding is required to fit the training circuit. The accepted security claim uses specified sieve cost models, not a universal security level; Core-SVP estimates are lower. Dense-secret hybrid search is bounded rather than an exhaustive global optimization. An all-sparse main-key alternative did not meet the selected security criterion and is not used. KAT coverage is limited to the supported SEAL preset, with asymmetric comparison directions; it does not independently validate our bootstrap or larger hybrid preset. Standalone OpenFHE/Lattigo bootstrap precision parity remains absent. Random full-slot CPU precision checks do not prove a negligible failure probability for arbitrary training distributions or activations outside range.

\paragraph{Learning limits.}
The quartic activation's claimed accuracy benefit is unmeasured. The evaluated encrypted trainer omits weight decay and optional softmax/cross-entropy. The separate plaintext paper-setting reconstruction includes weight decay; it is not the faithful encrypted protocol. Bias momentum is not matched to NAG, and faithful bias freezing preserves extra encrypted arithmetic. Fixed-final-epoch Breast Cancer accuracy misses its target; test-selected best epochs are not a remedy. No encrypted Fashion-MNIST or Breast Cancer accuracy is reported. Higher-depth local-block training has plaintext fidelity checks but no completed encrypted comparison. The continuity Iris experiment is interactive and differs in padding, randomization, and normalization; it cannot establish the primary non-interactive claim. Threshold or multi-party FHE, scheme switching, transformer training, and Metal execution are outside the evaluated scope.

\paragraph{Device and implementation limits.}
The executed full-slot bootstrap gate and short main-model CPU/CUDA pair establish their measured numerical scopes. The matched pair extends exact decoded CPU/GPU agreement to 100 updates. The completed ten-pass GPU run measures longer-horizon accuracy and accumulated error, but historical CPU FULL-2 has a different executable and no exported weights; exact full-run parameter parity remains unverified. Both studies use one initialization. Untested full-key operator cases remain unverified. Repaired multi-stream latency is measured only for independent products at a capped level; it cannot substitute for training measurements. Memory estimates and pool snapshots do not establish a peak. All-off flags do not implement the strongest naive diagonal algorithm. The benchmark supports device bootstrap explicitly and retains the host adapter for reproduction. Device error handling and numerical replay are useful engineering evidence, but compile-only paths and skipped tests are not physical hardware results. CPU/GPU side-channel resistance, constant-time sampling, secret zeroization, malformed-input hardening, and independent cryptographic audit are outside the current implementation's assurance. Some internal invariants are debug assertions; decode/parameter mismatches can panic. The artifact is research software, not a deployment security claim.

\paragraph{Model extensions.}
Wider MLPs provide a controlled test of the layout's scaling, but do not by themselves establish architectural novelty: ReBoot already evaluates multiple local-block architectures. A subsequent research direction is convolutional training with encrypted updates to shared kernels. This requires jointly planning the forward pass, kernel and input gradients, optimizer state, and refresh schedule; implementing convolutional inference alone would not validate that training path. We would first retain the present polynomial activation to isolate the effects of packing and memory management. These CNN extensions are not evaluated in this work.

\subsection{Conclusion}
Batched diagonal packing preserves a rotation-only transpose while supporting compact non-interactive optimizer updates. The completed full-setting runs show a larger-batch count advantage and baseline-batch accuracy within tolerance, together with a missed normalized count target at the baseline batch. The security mismatch and unmeasured full-setting baseline time prevent an unconditional superiority claim. Device-resident bootstrap removes the dominant host-refresh bottleneck in our CUDA training path. A matched 100-update experiment yields a $48.68\times$ training-time ratio against our CPU implementation with exactly equal decoded parameters and outputs ($40.67\times$ including setup and final evaluation). A separate ten-pass GPU run completes 1,000 updates in 2,967.095 seconds with 90.35\% test accuracy, 0.06 percentage points below its plaintext twin. Full-run parameter parity and multi-seed encrypted behavior remain unverified. These measurements strengthen the implementation evidence while preserving the unmet matched-security and full-setting ReBoot timing conditions.

\bibliographystyle{ACM-Reference-Format}
\bibliography{refs}


\begin{thebibliography}{13}


\ifx \showCODEN    \undefined \def \showCODEN     #1{\unskip}     \fi
\ifx \showISBNx    \undefined \def \showISBNx     #1{\unskip}     \fi
\ifx \showISBNxiii \undefined \def \showISBNxiii  #1{\unskip}     \fi
\ifx \showISSN     \undefined \def \showISSN      #1{\unskip}     \fi
\ifx \showLCCN     \undefined \def \showLCCN      #1{\unskip}     \fi
\ifx \shownote     \undefined \def \shownote      #1{#1}          \fi
\ifx \showarticletitle \undefined \def \showarticletitle #1{#1}   \fi
\ifx \showURL      \undefined \def \showURL       {\relax}        \fi
\providecommand\bibfield[2]{#2}
\providecommand\bibinfo[2]{#2}
\providecommand\natexlab[1]{#1}
\providecommand\showeprint[2][]{arXiv:#2}
\makeatletter
\@ifundefined{NAT@parse@date}{}{\let\NAT@parse@date@orig\NAT@parse@date}
\@ifundefined{NAT@parse@date}{}{\def\NAT@parse@date#1#2#3#4#5#6@@{\NAT@parse@date@orig#1#2#3#4#5#6@@\def\NAT@tempyear{0000}\def\NAT@tempexlab{{?}}\ifx\NAT@year\NAT@tempyear\ifx\NAT@exlab\NAT@tempexlab\def\NAT@date{[n.\,d.]}\else\edef\NAT@date{[n.\,d.]\NAT@exlab}\fi\fi}}
\makeatother

\bibitem[{Al Badawi} et~al\mbox{.}(2022)]%
        {openfhe2022}
\bibfield{author}{\bibinfo{person}{Ahmad {Al Badawi}}, \bibinfo{person}{Jack Bates}, \bibinfo{person}{Flavio Bergamaschi}, \bibinfo{person}{David~Bruce Cousins}, \bibinfo{person}{Saroja Erabelli}, \bibinfo{person}{Nicholas Genise}, \bibinfo{person}{Shai Halevi}, \bibinfo{person}{Hamish Hunt}, \bibinfo{person}{Andrey Kim}, \bibinfo{person}{Yongwoo Lee}, \bibinfo{person}{Zeyu Liu}, \bibinfo{person}{Daniele Micciancio}, \bibinfo{person}{Ian Quah}, \bibinfo{person}{Yuriy Polyakov}, \bibinfo{person}{Saraswathy~R. V.}, \bibinfo{person}{Kurt Rohloff}, \bibinfo{person}{Jonathan Saylor}, \bibinfo{person}{Dmitriy Suponitsky}, \bibinfo{person}{Matthew Triplett}, \bibinfo{person}{Vinod Vaikuntanathan}, {and} \bibinfo{person}{Vincent Zucca}.} \bibinfo{year}{2022}\natexlab{}.
\newblock \showarticletitle{{OpenFHE}: Open-Source Fully Homomorphic Encryption Library}. In \bibinfo{booktitle}{\emph{Proceedings of the 10th Workshop on Encrypted Computing \& Applied Homomorphic Cryptography}}. \bibinfo{pages}{53--63}.
\newblock
\href{https://doi.org/10.1145/3560827.3563379}{doi:\nolinkurl{10.1145/3560827.3563379}}


\bibitem[Albrecht et~al\mbox{.}(2015)]%
        {albrecht2015}
\bibfield{author}{\bibinfo{person}{Martin~R. Albrecht}, \bibinfo{person}{Rachel Player}, {and} \bibinfo{person}{Sam Scott}.} \bibinfo{year}{2015}\natexlab{}.
\newblock \showarticletitle{On the concrete hardness of Learning with Errors}.
\newblock \bibinfo{journal}{\emph{Journal of Mathematical Cryptology}} \bibinfo{volume}{9}, \bibinfo{number}{3} (\bibinfo{year}{2015}), \bibinfo{pages}{169--203}.
\newblock
\href{https://doi.org/10.1515/jmc-2015-0016}{doi:\nolinkurl{10.1515/jmc-2015-0016}}


\bibitem[Bossuat et~al\mbox{.}(2022)]%
        {sse2022}
\bibfield{author}{\bibinfo{person}{Jean-Philippe Bossuat}, \bibinfo{person}{Juan~Ram{\'o}n Troncoso-Pastoriza}, {and} \bibinfo{person}{Jean-Pierre Hubaux}.} \bibinfo{year}{2022}\natexlab{}.
\newblock \bibinfo{title}{Bootstrapping for Approximate Homomorphic Encryption with Negligible Failure-Probability by Using Sparse-Secret Encapsulation}.
\newblock \bibinfo{howpublished}{Cryptology ePrint Archive, Paper 2022/024}.
\newblock
\urldef\tempurl%
\url{https://eprint.iacr.org/2022/024}
\showURL{%
\tempurl}


\bibitem[Cheon et~al\mbox{.}(2017)]%
        {ckks2017}
\bibfield{author}{\bibinfo{person}{Jung~Hee Cheon}, \bibinfo{person}{Andrey Kim}, \bibinfo{person}{Miran Kim}, {and} \bibinfo{person}{Yongsoo Song}.} \bibinfo{year}{2017}\natexlab{}.
\newblock \showarticletitle{Homomorphic Encryption for Arithmetic of Approximate Numbers}. In \bibinfo{booktitle}{\emph{Advances in Cryptology -- ASIACRYPT 2017}}. \bibinfo{pages}{409--437}.
\newblock
\href{https://doi.org/10.1007/978-3-319-70694-8_15}{doi:\nolinkurl{10.1007/978-3-319-70694-8_15}}


\bibitem[Halevi and Shoup(2018)]%
        {halevi2018}
\bibfield{author}{\bibinfo{person}{Shai Halevi} {and} \bibinfo{person}{Victor Shoup}.} \bibinfo{year}{2018}\natexlab{}.
\newblock \bibinfo{title}{Faster Homomorphic Linear Transformations in {HElib}}.
\newblock \bibinfo{howpublished}{Cryptology ePrint Archive, Paper 2018/244}.
\newblock
\urldef\tempurl%
\url{https://eprint.iacr.org/2018/244}
\showURL{%
\tempurl}


\bibitem[{HEonGPU Project}(2026)]%
        {heongpuBench}
\bibfield{author}{\bibinfo{person}{{HEonGPU Project}}.} \bibinfo{year}{2026}\natexlab{}.
\newblock \bibinfo{title}{{HEonGPU}: Appendix, Performance Benchmarks}.
\newblock
\urldef\tempurl%
\url{https://heongpu.readthedocs.io/en/latest/appendix.html}
\showURL{%
\tempurl}
\newblock
\shownote{Accessed 2026-11-06; published workload differs from this study}.


\bibitem[{Lattice Estimator Contributors}(2026)]%
        {latticeEstimator}
\bibfield{author}{\bibinfo{person}{{Lattice Estimator Contributors}}.} \bibinfo{year}{2026}\natexlab{}.
\newblock \bibinfo{title}{Security Estimates for Lattice Problems}.
\newblock
\urldef\tempurl%
\url{https://github.com/malb/lattice-estimator}
\showURL{%
\tempurl}
\newblock
\shownote{Documentation accessed 2026-11-06; experiment commit recorded in Appendix A}.


\bibitem[Lee et~al\mbox{.}(2023)]%
        {hetal2023}
\bibfield{author}{\bibinfo{person}{Seewoo Lee}, \bibinfo{person}{Garam Lee}, \bibinfo{person}{Jung~Woo Kim}, \bibinfo{person}{Junbum Shin}, {and} \bibinfo{person}{Mun-Kyu Lee}.} \bibinfo{year}{2023}\natexlab{}.
\newblock \showarticletitle{{HETAL}: Efficient Privacy-preserving Transfer Learning with Homomorphic Encryption}. In \bibinfo{booktitle}{\emph{Proceedings of the 40th International Conference on Machine Learning}} \emph{(\bibinfo{series}{Proceedings of Machine Learning Research}, Vol.~\bibinfo{volume}{202})}. \bibinfo{publisher}{PMLR}, \bibinfo{pages}{19010--19035}.
\newblock
\urldef\tempurl%
\url{https://proceedings.mlr.press/v202/lee23m.html}
\showURL{%
\tempurl}


\bibitem[{Microsoft Research}(2026)]%
        {seal}
\bibfield{author}{\bibinfo{person}{{Microsoft Research}}.} \bibinfo{year}{2026}\natexlab{}.
\newblock \bibinfo{title}{{Microsoft SEAL}}.
\newblock
\urldef\tempurl%
\url{https://github.com/microsoft/SEAL}
\showURL{%
\tempurl}
\newblock
\shownote{Project documentation accessed 2026-11-06; experiments use version 4.1.2}.


\bibitem[Mihara et~al\mbox{.}(2020)]%
        {mihara2020}
\bibfield{author}{\bibinfo{person}{Kentaro Mihara}, \bibinfo{person}{Ryohei Yamaguchi}, \bibinfo{person}{Miguel Mitsuishi}, {and} \bibinfo{person}{Yusuke Maruyama}.} \bibinfo{year}{2020}\natexlab{}.
\newblock \bibinfo{title}{Neural Network Training With Homomorphic Encryption}.
\newblock \bibinfo{howpublished}{arXiv:2012.13552}.
\newblock
\href{https://doi.org/10.48550/arXiv.2012.13552}{doi:\nolinkurl{10.48550/arXiv.2012.13552}}


\bibitem[{\"O}zcan and Sava{\c s}(2025)]%
        {gpuKeys2025}
\bibfield{author}{\bibinfo{person}{Ali~{\c S}ah {\"O}zcan} {and} \bibinfo{person}{Erkay Sava{\c s}}.} \bibinfo{year}{2025}\natexlab{}.
\newblock \bibinfo{title}{{GPU} Implementations of Three Different Key-Switching Methods for Homomorphic Encryption Schemes}.
\newblock \bibinfo{howpublished}{Cryptology ePrint Archive, Paper 2025/124}.
\newblock
\urldef\tempurl%
\url{https://eprint.iacr.org/2025/124}
\showURL{%
\tempurl}
\newblock
\shownote{Author-deposited manuscript dated November 2024; Table 3 checked 2026-11-08}.


\bibitem[Pirillo and Colombo(2026)]%
        {reboot2026}
\bibfield{author}{\bibinfo{person}{Alberto Pirillo} {and} \bibinfo{person}{Luca Colombo}.} \bibinfo{year}{2026}\natexlab{}.
\newblock \showarticletitle{{ReBoot}: Encrypted Training of Deep Neural Networks with {CKKS} Bootstrapping}.
\newblock \bibinfo{journal}{\emph{Proceedings of the AAAI Conference on Artificial Intelligence}} \bibinfo{volume}{40}, \bibinfo{number}{29} (\bibinfo{year}{2026}), \bibinfo{pages}{24837--24845}.
\newblock
\href{https://doi.org/10.1609/aaai.v40i29.39670}{doi:\nolinkurl{10.1609/aaai.v40i29.39670}}


\bibitem[{WAHC Organizers}(2026)]%
        {wahc2026}
\bibfield{author}{\bibinfo{person}{{WAHC Organizers}}.} \bibinfo{year}{2026}\natexlab{}.
\newblock \bibinfo{title}{{WAHC} 2026: Call for Papers}.
\newblock
\urldef\tempurl%
\url{https://homomorphicencryption.org/wahc-2026-14th-workshop-on-encrypted-computing-applied-homomorphic-cryptography/}
\showURL{%
\tempurl}
\newblock
\shownote{Accessed 2026-11-06; 12 pages including references and appendices; 2027 rules unconfirmed}.


\end{thebibliography}
\appendix
\section{Parameters, Security, and Reproduction}
\label{app:params}
\subsection{Implemented chain and bootstrap contract}
The production bootstrap preset uses $N=2^{16}$, $N/2$ real slots, scale $2^{42}$, and Q-prime bit sizes
\begin{equation*}
 [54,\underbrace{42,\ldots,42}_{16},\underbrace{34,\ldots,34}_{4},
 \underbrace{58,\ldots,58}_{7},\underbrace{56,\ldots,56}_{4}].
\end{equation*}
Four special primes have bit size $60$, with eight hybrid digits. The bands allocate four levels to each linear transform and seven to EvalMod, leaving level $16$ before training unpack. EvalMod uses Chebyshev degree $127$, interval parameter $K=13$, and no double-angle step. These are implemented design constants; the superseded original-plan bands and dense-secret range are not used.

The bootstrap requires level zero and exactly the preset scale before work. Dense-to-sparse SSE is restricted to $q_0P$ under a temporary key of weight $32$; the return key and remaining arithmetic use the dense main key over divisors of $QP$. CPU full-slot tests exceed the registered precision threshold for random real inputs in $[-1,1]$. This does not imply a worst-case failure bound. The executed RTX~4090 full-slot gate matches CPU limbs and counters and measures 17.4775 precision bits. The completed main-model GPU pass establishes feasibility for that workload; pool checkpoints do not establish peak memory or feasibility for larger models.

\subsection{What the estimator models}
Our default sampler gives $-1,0,1$ probabilities $1/4,1/2,1/4$. We model the dense secret with a fixed-count sparse-ternary proxy rather than a uniform-ternary distribution; SSE is a separate small-modulus instance. The reported attack input rounds the exposed $\log_2(QP)$ upward and uses $m=N$ samples. Dense and sparse main keys must not be interchanged at the same modulus without re-estimation.

The estimator is pinned to commit \texttt{ae45d740ceb1}; methodology follows Albrecht et al.~\cite{albrecht2015} and the maintained estimator documentation~\cite{latticeEstimator}. The recorded search includes primal/dual and selected hybrid/MITM variants. Dense primal-hybrid optimization did not finish globally; fixed guessing-dimension scans supplement it. Arora--GB and BKW were not included. For the released ReBoot context, an observed sign split from one generated key is an input instance, not a distributional guarantee for all keys. The minima are conditional on the selected search, cost model, and memory assumptions. Table~\ref{tab:security} is not a proof, nor evidence of a practical compromise.

\subsection{Transpose identity and reduction boundary}
For Eq.~\eqref{eq:transpose}, matching row and column indices gives $t\equiv i\pmod {n_o}$ and $t+c\equiv0\pmod {n_i}$. The wide/tall cases in Section~\ref{sec:packing} solve both conditions. Substituting the pre-rotations on the source and destination yields the displayed offset. Because every weight diagonal is periodic with period $L$ in feature coordinates and all offsets are multiples of $B$, samples are not mixed. The identity therefore applies to rectangular padded layers and repeated blocks as well as square matrices. A zero offset is a copy; other offsets require the selected dedicated or composed rotation keys.

Lane reduction is different: offsets not divisible by $B$ mix lanes intentionally, and isolating one lane requires a mask. Ignoring that mask or its replication work changes both the depth and count model. A partial-batch validity mask must also exclude unused lanes from the loss derivative; zero input alone is insufficient when the model has a bias or a linear output loss.

\subsection{KAT and artifact boundaries}
The SEAL compatibility checks compare encoding, serialized ciphertext interoperability, decryption, and supported evaluation paths under the shared preset. Some directions assert exact residues while others only assert a decoded-error bound. Rotation arithmetic intentionally differs in its centered lifting. Neither this KAT nor CPU/GPU bit equality establishes security; the latter only checks implementation agreement for executed cases. Small insecure presets are smoke tests and never enter the security claim.

The comparison-table generator exposes field-level provenance through its \texttt{--sources} option. A separate device-training validator generates its table after checking captured original results, executable and input hashes, counts, timing, and numerical agreement. The figure generator consumes those validated table cells and writes a source manifest containing paths, fields, and SHA-256 digests. It does not rewrite result JSON. Missing registered runs produce explicit pending rows; existing invalid inputs must fail validation. Timed runs, setup, client diagnostics, and model-only extrapolations stay distinguishable. Running \texttt{scripts/run\_all.sh --tables-only} regenerates the retained tables. This does not execute the complete registered experiment matrix, whose end-to-end runner remains unfinished; a quick pipeline check is not a source of research results.

The anonymous review mode omits author metadata and identifying artifact links. ReBoot and OpenFHE are executed as external baselines; no implementation source is copied into this artifact. The estimator and TeX compiler are isolated development tools, not library dependencies. The library's dependency policy, including the specific character-table exception disclosed in Section~\ref{sec:limits}, is separate from the licenses of those external research tools. Instructions for rebuilding anonymous review, named review, and named preprint variants are provided with the source. Authorship is confirmed for this preprint. Before arXiv submission, the final source, metadata, and server-generated PDF require checking against the retained evidence; unavailable measurements remain unavailable. A later conference submission requires a separate anonymity and venue-rules review; the WAHC~2026 limit is only the provisional formatting constraint~\cite{wahc2026}.

\ifdefined\ArxivDraft
Code and evidence: \url{https://github.com/kenmaro3/ckks-train}. Release \texttt{v1.0.0-preview.1} is an earlier snapshot; this preprint has its own source revision.
\fi

\end{document}